\documentstyle[epsfig,prc,aps]{revtex}
\begin{document}

\title{ $ \pi \Lambda \Sigma  $ Coupling Extracted from Hyperonic Atoms}

\author{B. Loiseau\thanks{{\it Internet address:} loiseau@in2p3.fr}}
\address{Laboratoire de Physique Nucl\'eaire et de Hautes
\'Energies\thanks{Unit\'e de Recherche des Universit\'es Paris 6 et Paris 7,
associ\'ee au CNRS},\\
LPTPE, 
Universit\'{e} P. \& M. Curie, 4 Place Jussieu, 75252 Paris Cedex 05, 
France}
\author{S. Wycech\thanks{{\it Internet address:} wycech@fuw.edu.pl}}
\address{So{\l}tan Institute for Nuclear Studies,
Ho\.za 69, PL-00-681, Warsaw, Poland}

\maketitle

\begin{abstract}
The latest measurements of the atomic level width in $\Sigma$-hyperonic
Pb atom offer the most accurate datum in the region of low-energy
$\Sigma$-hyperon physics. Atomic widths are due to the
$\Sigma N-\Lambda N$ conversion which in high angular momentum
states is dominated by the one pion exchange. A joint analysis of the
$ \Sigma^- p \rightarrow  \Lambda n $  scattering data and atomic widths
allows to extract the pion-hyperon  coupling constant
$ f_{\pi \Lambda \Sigma}^{2}/4\pi = 0.048 \pm 0.005$ (statistic)
$\pm 0.004$ (systematic) or $G^2_{\pi \Lambda \Sigma} = 13.3 \pm 1.4$
(statistic) $\pm 1.1$ (systematic).
\end{abstract}

\section{ Introduction}

It has been known  for some time that hadronic atoms 
allow to test the long range part of hadron-nucleon interactions, 
the one-pion exchange (OPE),~\cite{pil,lan}.
Ideal conditions are found in high angular momentum atomic states. 
There, the centrifugal barrier prevents the hadron  to get close 
to the nucleus and the OPE effect is strongly enhanced over the 
effects of two-pion or heavier meson exchanges. 
Detailed investigations are rather involved. In the best studied
cases of antiprotonic atoms most of the long range pionic 
effects disappear due to spin and isospin averaging over  
the nuclear states of the bound nucleon. 
Decays of  $\Sigma$-hyperonic atoms offer here a better chance, 
even though the experimental data  are in general rather uncertain 
\cite{bac,bat78,pow}. One exceptional case is the relatively recent 
measurement of Powers {\it et al.}~\cite{pow} that provides a  
precise  width of 17(3)~eV for the $\Sigma^{-}$Pb atom in an 
"upper" $n=10, l =$ 9 state. It gives a good opportunity to study 
the role of one-pion exchange. 

The lifetimes of  $\Sigma$ atoms are  determined by the rate
of the reaction 
\begin{equation}
\Sigma^- + p \rightarrow  \Lambda + n  
\label{1}
\end{equation}
occurring at the nuclear surface. This process of hyperon 
conversion is dominated by the OPE and we exploit this feature 
to do a best estimate of the $f_{\pi \Lambda \Sigma}^{2} $ coupling constant. 
Reaction (\ref{1}) has been studied in  scattering experiments 
at low energies but the data are sparse and not very precise 
\cite{eis,ste}. Differential cross sections are very 
uncertain. These, as well as the elastic and other inelastic 
hyperon interactions, have been described by  meson-exchange  
models  \cite{swa,sto,rij,reu}. In addition to OPE, other short-range
 interactions are found to contribute. On the other hand, 
in high angular momentum atomic states of  $\Sigma$ atoms, the 
effects of short-range forces are strongly suppressed. 
The $\Sigma-N $ interactions in these atomic states 
are somehow equivalent to interactions in high partial waves of the free   
$\Sigma-N $ system. However, there is a price for this simplicity: 
the atomic 
system involves the rather intricate nuclear structure. 
The latter  determines the initial state of the atom  and affects the 
final decay states. Fortunately,  the conversion 
happens on a proton and the nuclear proton-density distributions are well 
known from $\mu$ atoms and  electron scattering data. In addition, 
the proton momenta 
are quite well determined by nuclear shell models. The final states
of the nucleus, $\Lambda$ and neutron
are less certain, however. These enter into the formalism, essentially  
in terms of the energy conservation that determines the energy and 
momentum transferred by the exchanged pion. In this paper it is shown that
the atomic-level widths are very stable against these 
uncertainties.   

In section II of this paper
we present the formalism and the determination of the $\pi \Lambda \Sigma $ 
coupling constant from the atomic widths. In addition to the Pb
datum, the most accurate older  results are also included into 
the best fit procedure. Some details of the OPE are given in 
Appendix A while Appendix B discusses the atomic wave functions. 
In section III limitations and uncertainties of our method
are discussed and the error of the extracted 
$\pi \Lambda \Sigma $ coupling constant is
evaluated. These limitations stem from two sources: first the 
hyperon conversion process is not  described by  the pion exchange 
alone, there exist other amplitudes due to short ranged processes.
Several models of these additional amplitudes are used and tested  
against the scattering data.
The effect of such amplitudes on the atomic widths is found to be
a $10\% $ correction. The determination  of the $\pi \Lambda \Sigma$ 
coupling is weakly affected. Other uncertainties are due to the 
nuclear physics involved in the process. All these generate 
systematic errors which are comparable to the statistical one. 
 
\section { The nuclear conversion  of atomic $\Sigma $ hyperons}

The $\Sigma^{-} $ hyperon,  bound into atomic orbits, cascades
down to  states where it becomes converted into $\Lambda $.
The quantum numbers and lifetimes of such states  have been
determined in  X-ray experiments \cite{bac},\cite{bat78},\cite{pow}.
It turns out that these are  levels of high angular momenta  which
force the conversion process to occur at the nuclear surface.
The overlaps of the nuclear and atomic densities are localized
in extreme surface regions and extend to distances as large as
twice the nuclear radius \cite{bat97}. Nuclear densities
involved in the "upper" levels (states of the highest available
angular momenta) amount to  $3\%$ of the central density.
Processes of such peripherality  allow for the standard low
density simplifications: quasi-free scattering and
single-particle picture  of the nucleus. The final states may
be described as free  states. On the other hand, there exist
a difficulty inherent to the surface studies related to the
sensitivity to all
ranges. For the problem
of interest, we turn it into an advantage as it will allow us to precise
the $\pi \Lambda \Sigma$ coupling.

This section presents the description of the $\Sigma - \Lambda$ 
conversion mechanism. It follows the calculation performed  
in Ref.\cite{wyc} for the antiprotonic atoms.
First, a simple phenomenological picture based on the optical  
potential model is presented. Next, we elaborate a unitarity 
relation  for reaction (\ref{1}) and finally some technical  
details of this calculation are discussed. 

The tool to describe the atomic-level shifts and widths due 
to nuclear   interactions is an optical potential, $V^{opt}$.  
The simplest one is  usually \cite{bat97} assumed to have the form
\begin{equation}
\label{1a}
      V^{opt}({\bf R})=\frac{2\pi}{\mu_{\Sigma N}} 
      \int d{\bf u} \ \rho({\bf R}-{\bf u}) \ t_{\Sigma N}({\bf u})
\end{equation}
where $\mu_{\Sigma N}$ is the $ \Sigma N $ reduced mass, ${\rho({\bf R})}$ 
is the nuclear 
density at a radius ${\bf R}$ and $t_{ \Sigma N}$ is the elastic scattering 
matrix. In Eq.(\ref{1a}) the finite range in the ${\Sigma N}$ 
interactions induces  a folding of the nuclear density $\rho $  
over this  range.   The atomic level widths  are  given by 
\begin{eqnarray}
\label{2a}
  \Gamma/2 & = & - \int d{\bf R}\ Im V^{opt}({\bf R})
  \mid\Psi_{ \Sigma}({\bf R})\mid^{2} \nonumber \\
 & = & -\frac{2\pi}{\mu_{\Sigma N}}\int d{\bf R}\int d{\bf u}\ 
  \rho({\bf R}-{\bf u}) 
 \ Im \ t_{\Sigma N}({\bf u}) \mid\Psi_{ \Sigma}({\bf R})\mid^{2}  
\end{eqnarray}
where $\Psi_{ \Sigma}({\bf R})$ is the atomic wave function that 
involves effects of nuclear interactions.
Since $\Psi_{ \Sigma} \sim R^l$  and since the angular momenta  
$l$ are large, the nuclear and atomic densities entering into 
Eq.(\ref{2a}) are well separated and it is only the longest 
range components of Im~$ t_{\Sigma N}({\bf u})$ that matter. To 
simplify the discussion we consider the widths of "upper" 
atomic levels, only. There are three advantages of this choice : 
i) the most precise data, ii) the largest $l$ and iii) the almost
unperturbed hydrogen-like atomic wave functions. The optical  
potential given by Eq.(\ref{1a}) has some phenomenological 
success \cite{bat97} but for the purpose of this study we need  
a more microscopic approach that discloses the strength and range  
involved in   Im~$t_{\Sigma N }$. This quantity 
is related  by unitarity to the inelastic reactions   
$ \Sigma^- p \rightarrow  \Lambda n ,\ \Sigma^{0}n $. In addition some 
specific atomic conditions must be met : the  initial hyperon 
and nucleon are bound and then the $\Sigma^{0} n $ channel is 
blocked~\cite{pil}. The unitarity condition is also affected by the 
presence of the nucleus. The latter takes away some recoil  
energy and may induce final state interactions of the outgoing 
hadrons. These interactions are neglected since the process is 
an extremely peripheral one, there are no Coulomb interactions 
and the final particles share about 70~MeV kinetic energy. On the other
hand an assessment of the nucleon binding, of the nuclear recoil and 
of the rearrangement energies is of  significance and will be studied 
below. 

\subsection{ The atomic-level widths and the unitarity condition }

The aim of this section is to calculate the rate of the nuclear 
$ \Sigma - \Lambda$ conversion. This is done in two steps: 

1) An amplitude for the $ \Sigma p  -  \Lambda n $ conversion    
is given in the leading order by OPE. This amplitude is 
introduced into the nuclear transition matrix element.   
Other  "background"  amplitudes, $t_{b}$, will be discussed later. 

2) The $\Lambda$ emission probabilities are calculated and  
summed over states of $ \Lambda$, $n $  and over states of the 
final nucleus. For an isolated ${\Sigma}\ p$ pair this procedure 
would produce the inelastic cross section and, via the unitarity 
condition, the absorptive amplitude Im $t_{ \Sigma N}$. Here, 
the summation over final states extends over free $\Lambda$,  
$ n $ states and over single proton-hole nuclear states. This 
summation generates Im $t_{\Sigma N}$ and in the nuclear case it is folded 
over nuclear and atomic wave functions. These  wave functions 
arise in the form of mixed single particle density matrices. 

Assume that the $\Sigma$ hyperon in an $n$-th atomic state 
of quantum numbers $[n,l,m]$ and a proton in a single particle 
state $\alpha$ convert into a free $\Lambda$ and a neutron.  
The transition amplitude for this process is
\begin{equation}
\label{5}
    A_{n,\alpha,P,q}=\int d{\bf x} d{\bf y}\ 
	\Psi_{ \Sigma}({\bf x}) \varphi^{\alpha}_N({\bf y})
	\ V_{ \pi}({\bf x}-{\bf y})\ \exp(i{\bf P}.{\bf R}+i {\bf q}.{\bf r})  
\end{equation}                        
where $ \varphi^{\alpha}_N({\bf y})$ are the nuclear wave 
functions and $V_{ \pi}({\bf x}-{\bf y}) $ is the OPE potential the expression
of which is derived in Appendix A. The momenta and coordinates 
(${\bf P},{\bf R}$) refer to the center of mass while 
(${\bf q},{\bf r}$) denote the relative coordinates  for 
the final pair.
The final nucleus is specified, in the spirit 
of the impulse approximation, to be the initial nucleus  
left with a hole in the single particle state ${\alpha}$. 

To calculate the absorption widths, amplitudes $ A_{n,\alpha,P,q}$
of Eq.(\ref{5}) are squared, summed over the final nuclear states  
and integrated over the phase space of the $\Lambda$,$n$ pair.
In this way one obtains the following expression for the atomic level widths: 

\begin{eqnarray}
\label{6}
\Gamma /2 =  \sum_{\alpha}  \int d({\bf RR'rr'})\ d({\bf Pq})&&
		\overline{\Psi}_{ \Sigma}({\bf R'}+a{\bf r'})
		 \ \Psi_{ \Sigma}({\bf R}+a{\bf r})\
		\overline{\varphi}^{\alpha}_{N}({\bf R'}-b{\bf r'})
	 \ \varphi^{\alpha}_{N}({\bf R}-b{\bf r}) \nonumber \\
       & & \times V_{\pi}({\bf r})V_{\pi}({\bf r'})\ 
	  \frac{\mu_{N\Lambda}\ \pi\ \delta (Q-E(P)-E(q))}{2\pi (2\pi)^{6}}    
	   \  \exp(i{\bf P}.({\bf R}-{\bf R'})+i {\bf q}.({\bf r}-{\bf r'})).
\end{eqnarray}
The coefficients $a = M_N /(M_N+M_{\Sigma}) $ and
$ b=M_{\Sigma}/(M_N+M_{\Sigma})$ come from the transformation to the
 C.M. coordinates.  The  energy excess
$ Q= M_{\Sigma}-M_{\Lambda}-E_B(\alpha)$ is due to the hyperon mass
difference reduced by the nucleon binding $ E_B(\alpha)$. The values
of $Q$ fall in the range of $70 - 80$~MeV, sizable  by the nuclear 
standards and lead to  simplifications that are valid at the nuclear 
surface. With  typical $\Sigma$ and $p$  momenta that are met in 
this region one finds the final C.M. recoil energy E(P) to be about 
10\% of the relative kinetic energy E(q). Hence an average value for this 
recoil energy is used in Eq.(\ref{6}). This  allows to do the 
integration over ${\bf P}$ which  generates a 
$ \delta({\bf R}-{\bf R'}) $ and brings some simplicity to the 
otherwise involved integral. The  calculations that follow 
are straightforward  although the numerics has to be performed with care.

The energy conservation fixes the final state relative momentum 
$q$ which is also the momentum carried by the $\pi $ meson. 
The actual value of $q$ depends on the nucleon binding and 
recoil energies. Its central value is about $260$~MeV  but 
some 10\%   changes follow from the distribution of 
$E_B(\alpha)+ E(P)$  which characterizes proton states localized
at the extreme nuclear surface.  
These changes are significant.  

\subsection {Range effects} 

In the limit of zero range $ \Sigma N$ interactions,
formula (\ref{6}) for the conversion width may be expressed 
in terms of nuclear densities. For finite-range interactions the 
single-particle wave functions generate mixed densities. However, for 
simplicity and  reference to the experimentally tested charge 
densities, one wants  an expression in terms of the true densities. 
At the nuclear surface this can be done with a good precision, 
at least for the  absorption  rates summed over all nucleon states. 
The relation that allows it has been obtained in Ref.\cite{neg} and reads
\begin{equation}
  \sum_{\alpha} \overline{\varphi}_N^{\alpha}({\bf R}-b{\bf r'})
				 \ \varphi_N^{\alpha}({\bf R}-b{\bf r})
  \approx \rho({\bf R} -b{\bf u})\ J_{nuc}(\tilde k\ bw)
\label{14}
\end{equation}
where   ${\bf w}={\bf r}-{\bf r'}$, ${\bf u}=({\bf r}+{\bf r'})/2$, 
$\rho({\bf R} -b{\bf u})$ is the proton density. The $J_{nuc}(x)=3j_{1}(x)/x$
is an analog of Wigner function which describes the proton
momentum distribution within the nucleus. The $\tilde k$ is
an effective local momentum.
It is calculable in a shell model and at the nuclear surface it 
can be related to the proton separation energy \cite{cam}.
In a similar way we express the angular averaged atomic wave functions 
$\overline{\Psi} \Psi$  by
\begin{equation}
\label{15}
\frac{1}{2l+1}\sum_{m}\overline{\Psi}_{ \Sigma}({\bf R}+a{\bf r'})
 \ \Psi_{ \Sigma}({\bf R}+a{\bf r}) \approx
\mid\Psi_{ \Sigma}({\bf R}+a {\bf u})\mid^{2} J_{at}(aw),
\end{equation}
and $J_{at}(aw)$ is obtained by an expansion of the atomic wave functions
in terms of $a {\bf w}$ (see Appendix B).
Collecting all formulas from (\ref{6}) to (\ref{15})
one arrives at an expression 
\begin{eqnarray}
\label{18}
  \Gamma /2 &=& N_f \frac {(\sqrt{2} f_{\pi N N} f_{\pi \Lambda \Sigma })^{2}} 
		{(m_{\pi}^{2} 4\pi)^{2} }\  
		\frac{\mu_{N\Lambda}}{2\pi}\ 
q\ \left (\frac{1}{3}\ m_{\pi}^{*4} + \frac{2}{3}\ q^{*4} \right) \nonumber \\
		& & \times \int d{\bf R} d{\bf u} 
		\mid\Psi_{ \Sigma}({\bf R}+a{\bf u})\mid^{2}
                \rho({\bf R}-b{\bf u})\ F_{fold}(u).
\end{eqnarray}  
The function $ F_{fold}$ that describes the range of atomic-nuclear 
folding is given by 

\begin{eqnarray}
\label{Ffold}
 F_{fold}(u)=  \int  d{\bf w} \frac{d \Omega_q}{4 \pi }&&
\frac{ \exp(-m^{*}\mid {\bf u}+{\bf w}/2 \mid)}
    { \mid {\bf u}+{\bf w}/2 \mid }\ 
 \frac{ \exp(-m^{*}\mid {\bf u}-{\bf w}/2 \mid)}
{ \mid {\bf u}-{\bf w}/2 \mid } \nonumber \\
&& \times   \exp(i{\bf q}.{\bf w})\ J_{nuc}(bw)\ J_{at}(aw).
\end{eqnarray}  

To obtain  the result of Eq.~(\ref{18}),
few simple manipulations were carried out :

1) As shown in Appendix A, the charged pion exchange involves some
energy transfer that modifies the pion mass $m_\pi$ into $m^*_\pi$. The
$N_f=1.02$ factor describes 
off-shell effects
for the bound nucleon. 
The pion exchange potential of Eq.(\ref{a2}),
corrected for the $\delta({\bf r})$ singularity by Eq.(\ref{a3}) has been
used. The derivative operators are transformed by partial integration to
derivatives over the initial and final state wave functions. These
derivatives contain one large term due to the final $\Lambda n$ relative
momentum ${\bf q} $ and smaller contributions due to relative momenta of
$\Sigma$ and $p$.  The large momentum produces a dominant $q^4$ term in the
square bracket of Eq.(\ref{18}). The effect of the initial momenta is
obtained by calculating the derivatives over nuclear densities, atomic
densities and correlation functions $J_{nuc}$~(\ref{14}) and
$J_{at}$~(\ref{15}). To the leading order
in $u/R,\ w/R$ these induce a change of $q^2$ into,
   $$q^{*2}=q^2 + [ a(l/R-1/Bn)+b/(2a_p)f_p]^2 -(\tilde k b/5)^2,$$
   implemented in Eq.(\ref{18}). It amounts to a 25\% correction, 
   to obtain it we used the atomic wave function of Eq.(\ref{B2}) and 
   a two parameter Fermi (2pF) nuclear density profile 
      $1/(1+e)$, where $e = exp[(c-R)/a_p]$, $c $ is the 
   half  density radius and $a_p$ is the surface thickness.
   In terms of the    latter $f_p = e/(1+e)$. 
   Other corrections due to $J_{at}$ and  terms $(u/R)^2,(w/R)^2 $ are 
   found to be negligible. 
   
2) The spins of initial and intermediate baryons are averaged and summed. 
   These generate the 1/3 and 2/3 weights in the square-bracket term 
   of Eq.(\ref{18}).

Before performing the numerical calculations of Eq.(\ref{Ffold}), let us
elucidate  the range involved in
the folding function $ F_{fold}$. For the sake of argument
let  $J_{nuc}=J_{at}=1$ for these  slowly varying functions.
By going to Fourier transforms one obtains:

\begin{equation}
\label{20}
  F_{fold}(u)=\frac{16}{\pi} \int  d{\bf p}\ \frac{d \Omega_q}{4 \pi }\
   \frac{ \exp(i2 {\bf u}.{\bf p})}{ (m^{*2}+({\bf p}-{\bf q})^2)
   \ (m'^2+({\bf p}+{\bf q})^2) } 
 \end{equation}
where $ m'$ should be equal to $m^{*}$ but we have allowed for a more
general situation which is met when several mesons  of different
masses are exchanged.  From Eq.(\ref{20}) one infers that if
$m^{*}=m'$ the integrand dependence on  the $ ({\bf p}.{\bf q})$ 
angle is rather moderate. The main  bulk of $F_{fold}(u)$ is given by
$ 16 \pi\ \exp\left(-2u\sqrt{m^{*2}+q^2}\right)/ \sqrt{m^{*2}+q^2}$. Its
range is related both to the mass of the exchanged  meson and to the 
momentum transfer, as expected from the uncertainty principle. 
More detailed calculations indicate some longer range contributions 
that induce  small oscillations for very large $u$.
For $ m' > m^{*} $  formula  (\ref{20}) generates an oscillatory
behavior of $F_{fold}$. This may be seen in the limit of large 
$m'$  which yields 
\begin{equation}
F_{fold} =16 \pi \ \frac{\sin(2qu)} {2qu}\ \frac{\exp(-2m^{*}u)} {m'^{2} u}.
\label{Ffoldlargemprime}
\end{equation}
Rapid oscillations in $F_{fold}$ at large values of $u$ 
suppress such  contributions  to the atomic width. 
The $ m' \gg m^{*}$ terms are due to interferences of 
the pion exchange amplitude and other amplitudes 
of shorter ranges in space. These may arise from :  
exchange of heavier mesons, multiple scattering effects that 
are supposed to be of ranges shorter  than $1/2m_{\pi}$ 
and  vertex functions as shown in Eq.(\ref{a4}).  

Now, we  attempt to describe the atomic level widths 
with the simple OPE mechanism including the form factor of Eq.~(\ref{a4}). 
The most  precise X-ray  measurements of the upper level widths are 
collected in
Table I.  These results are reproduced by Eq.(\ref{18}) with the best fit 
value $ f_{\pi \Lambda \Sigma}^{2}/4 \pi = 0.050(6)$
on the confidence level of $\chi ^2_{/data}=2.38/4$. 
In fact the fitted parameter is $f_{\pi N N} f_{\pi \Lambda \Sigma}$,
and to obtain the pion-hyperon coupling  an average value of 
$f_{\pi N N} ^{2}/4\pi=0.08$  was used. 
These results were obtained with proton densities extracted from the 2pF 
charge densities of Fricke et.al ~\cite{dens}. The choice of these 
densities and the method to obtain the proton densities is discussed in 
more detail in the next section.
The error of $ f_{\pi \Lambda \Sigma}$ given above is due to 
the experimental uncertainties. 
In addition, there exist uncertainties of the nuclear structure 
and simplifications in the conversion mechanism which  induce 
systematic errors.  
To obtain these we analyze  the $ \Sigma p $ inelastic scattering 
data and discuss the basic nuclear parameters.

\section{Results}

The nuclear parameters of importance in this calculation are  
the energy excess $Q$ and the nuclear  recoil energy $ E(P)$.
The energy excess is determined  essentially by the separation 
energy of the valence protons. 
The recoil energy  is given by the $\Sigma p $ pair momentum 
distribution,  which follows the Fourier transforms of 
$\Psi_{ \Sigma}({\bf x})\times \varphi^{\alpha}_N({\bf x})$. These 
functions are localized around a point $R_o$ at the nuclear 
surface. Thus the Fourier transform is peaked at $P\approx L/R_o$ 
where $L$ is the angular momentum of the pair with respect 
to the nucleus. The $L$ is a sum  of the atomic and nuclear 
single particle angular momenta ${\bf L} ={\bf l}_a + {\bf l}_n $. 
Since $l_a > l_n$ the central value of $L$ is $l_a$ and the 
most likely momentum  is  $P = l_a/R_o \sim 1 fm ^{-1}$. This 
estimates the recoil 
energy to be about 10~MeV. More detailed calculations indicate 
the spread of the recoil energies of some 5~MeV half-width. 
As discussed in the previous section an average energy 
$ D= \langle E_B(\alpha)+ E(P)\rangle$ is used to  determine $q$, the 
momentum transferred by the pion,  and the  range involved in the  
folding function  $F_{fold}(u)$. The shell model generates rather 
small values for $D$ and  $ D/E(q) < 0.2$. Nevertheless, 
uncertainties may arise, due to the way one averages over the 
binding energies $ E_B(\alpha)$ or one calculates the recoil 
momenta $P$. These may be reflected in a few percent uncertainty 
in $q$. This problem is fortunately removed by a remarkable 
numerical stability  of the atomic level width 
with respect to changes of $ D$. This effect is related to the form 
of the pion coupling in Eq.(\ref{18}). For an increased (decreased) 
$q$ one finds a decrease (increase) in the atomic-nucleus overlap 
integral which is perfectly balanced by the increase
(decrease) of $q\times(q^{*4})$ in Eq.(\ref{18}). This property holds 
on a few percent level for D in the region of 5 to 25~MeV. In this calculation 
we use an average value of $ D[MeV]= E_s+ 10$   where $E_s$ is the proton 
separation energy. Changing D  
by $\pm$~5~MeV one obtains a change of 0.001 for the best fit 
$f_{\pi \Lambda \Sigma}^{2}$.  This change is used as a partial 
estimate of our  systematic error. 

\subsection{Sensitivity study to short range amplitudes}

So far the $\Sigma p - \Lambda n  $ conversion  has been described
by the long range OPE  amplitude. The justification of this
approximation  is the high angular momentum barrier that separates
the atomic hyperon and the nuclear proton.
An additional finding of the previous section is the suppression
of an interference of this pion amplitude with amplitudes related
to a shorter spatial extent (see Eq.~(\ref{Ffoldlargemprime}).
A check for these  additional "background" amplitudes  is possible  since
the free space conversion cross sections are known. In principle, one
needs a full interaction model to do that. However the data on
the $YN$ interactions are sparse and the models \cite{swa} - \cite{reu}
differ in their content and their results.
Here, we take a phenomenological attitude. A plausible
form of the background amplitude, motivated by these theoretical models,
is assumed. One or two free parameters are used to describe its magnitude.
Next
these strength parameters are fitted to reproduce the experimental conversion
cross sections. The effect of this "background" amplitude on the atomic
level width is calculated and the best fit to both atomic and
 scattering data is used to extract $ f_{\pi \Lambda \Sigma}^{2} $.
Several forms of the background amplitude, $t_b$ are used and the coupling 
constants obtained in this way are compared and "averaged". Such 
a procedure allows for an estimate of a systematic error in the coupling 
constant. 

The choice for additional amplitudes is based on several observations :

(1) On the ground of Lippmann--Schwinger equation one expects the 
scattering $t(r)$ matrix to be of the form 

\begin{equation}
\label{eq:tmatrix}
  t({\bf r}) = V({\bf r}) + V({\bf r}) \int d{\bf r'}\ G({\bf r}-{\bf r'})
  \ t({\bf r'}) 
  \end{equation}
where $t({\bf r})=t_\pi({\bf r})+t_b({\bf r})$.
 The second term of this equation is due to multiple scattering effects. 
 The "background" amplitude of interest may be due to the short range 
 part of potential $V(r)$ and/or to the multiple scattering contribution.
 What is significant is that the range involved in both these 
 terms is expected to be at least shorter than that of the two-pion exchange
 amplitude.

(2) The low momentum ($140 -170$~MeV/c)  inelastic $ \Sigma^{-} p$ cross
section   of Engelman \cite{eis} shows no structure and no enhancement
at low energies that would indicate a large scattering length in this
system. The K- matrix parameterization for a coupled
$\Lambda N,\ \Sigma N $ system in this state indicates the dominance of
inelastic processes \cite{swa}. 
However, to be on a safe side, we fit the Massachusetts data \cite{ste}
taken at higher momenta ($200 -580$~MeV/c, 8 data points)  to be 
more independent of the initial state interactions. 

\begin{figure}
\begin{center}
\vspace{13.0cm}
\epsfig{file=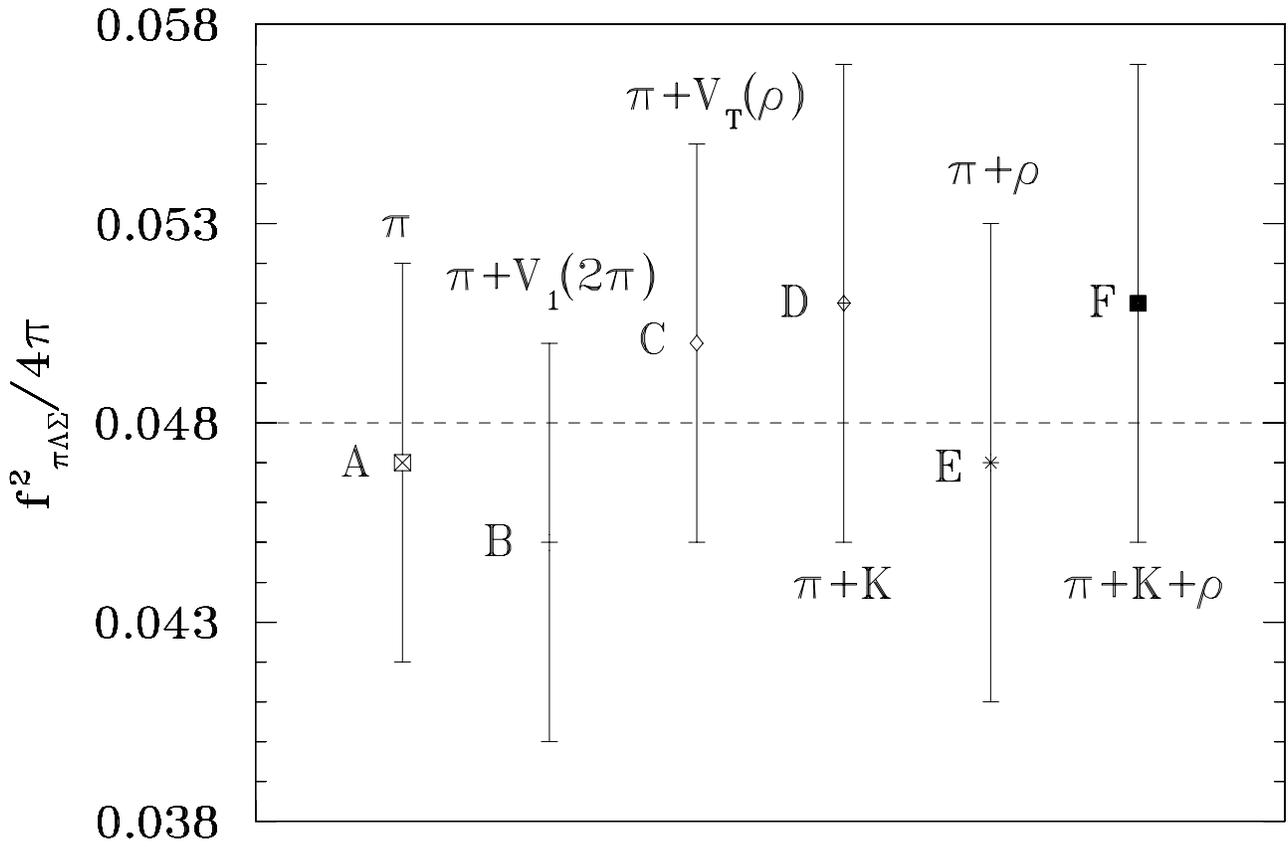,angle=90,height=11cm}
\vspace{0.5cm}
\caption{Results of the best fits to atomic and scattering data with
different background amplitudes (see text). Errors are the quadratic sum of
statistical and systematic uncertainties}
\label{fig:fpls}
\end{center}
\end{figure}

We consider  several models with different background amplitudes.
The results of the best fits to the atomic and scattering data
are summarized in Table II and plotted in Fig.~\ref{fig:fpls}.
The content of these models is as follows:

A) OPE approximation is used to describe both the atomic widths and
   the reaction cross sections. These quantities are dominated
   by the tensor part of OPE. The atomic data by itself favor 
   somewhat  stronger  $ f_{\pi \Lambda \Sigma}^{2}$ coupling, and 
   this is reflected in a small reduction of the best fit in 
   comparison to the result of the previous chapter . 
   The overall fit is very good, and this leaves very little room 
   for other contributions to the $\Sigma - \Lambda $ conversion.

B) The two pion effects are simulated by Yukawa potential of
   the $1/(2 m_{\pi})$ range and $f_{b}^{2}$ coupling. It is
   supplemented with a form factor of  source size parameter
   $\lambda~=~1100$~MeV. This force comes predominantly
   from a repetition of the tensorial force (see Eq.~\ref{eq:tmatrix})
   and contributes
   mostly to the spin triplet state, so the spin dependence of this
   potential is chosen to be
   (3 + {\boldmath $\sigma$}$_N$.{\boldmath $\sigma$}$_Y$)/4.
   The effect on the 
   best fit  $ f_{\pi \Lambda \Sigma}^{2}$ is minute. Similar 
   results are obtained with an assumption of a tensorial 
   form of this two pion force. The background  amplitudes 
   contribute some 10\% to the atomic widths. 
   A remark could be added at this point. Here one has a charged 
   particle exchange and one knows in the $NN$ sector that the 
   uncorrelated charged two-pion exchange gives a small
   contribution, there it is dominated by the $\rho$. 
   One could expect here a similar situation.

C) A tensor component of the force~(\ref{a1}) is used with a large $m'=m_\rho$ 
   mass of $770$~MeV and a coupling $f_{b}^{2}$. 

D) From SU(3) symmetry one expects a strong coupling with
   $K $ meson as found in Refs.~\cite{swa,reu}.
   The pseudoscalar exchange potential of the type~(\ref{a2}) is used
   with form factors  $\lambda = 1100$~MeV. Our  best fit coupling
   constant
   $f_{b}^{2}=\sqrt{2}f_{K N \Sigma}\ f_{K N \Lambda}=-0.19 $ to be
   compared to the  value $-$ 0.387 of Ref.~\cite{swa} and $-$ 0.416 of
   Ref.~\cite{reu}, although a much smaller number $-$ 0.030 is obtained in 
   Ref.~\cite{sto}. However, the "background"  constants obtained here 
   are just indicative. The data sample is not large enough for a real 
   determination. These couplings contribute very little to the 
   atomic widths and the corresponding $\chi^2$ minima  are 
   quite shallow. The "background" amplitudes are used only to 
   estimate the effects of  possible short range forces compared
   to the basic longest range OPE. 

E) The $\rho$  meson exchange amplitude of coupling
\begin{equation}                                          
\label{rho}
   (\mbox {\boldmath $\sigma$} \times {\bf q}).
   (\mbox {\boldmath $\sigma$} \times {\bf q})
  =\frac {1}{3} \left (-S_T\ q^2+2 \mbox {\boldmath $\sigma$}_N.
  \mbox {\boldmath $\sigma$}_Y \ q^2\right )
\Rightarrow  \frac {1}{3} \left (-S_T\ q^2-2 \mbox {\boldmath $\sigma$}_N.
\mbox {\boldmath $\sigma$}_Y\ m_\rho^{2}\right )
\end{equation}  
   is used with the mass  $m_\rho=770$~MeV and the cutoff range 
   $\lambda = 1100$~MeV.

F) The $ \rho+ K $  meson exchange amplitude. This is a three parameter 
   fit but only a marginal improvement has been obtained.  Again the 
   background amplitudes contribute about 10\%  to the dominant 
   OPE.  Now $ f_{K}^{2}= -0.20 $ and    $f_{\rho}^{2}= 1.7$.

These results show that the best fit value of $ f_{\pi \Lambda \Sigma}^{2}$
depends rather weakly on the uncertain short range interaction mechanism as
can be seen in Fig.~\ref{fig:fpls}. There the errors shown for each model
correspond to the quadratic sum of the statistical and systematic
uncertainties discussed previously. From Table II the systematic error due to
the imprecise knowledge of the short range interaction is found to be 0.003.

\subsection{ The choice of nuclear densities  }

Nuclear factors of prime importance are the proton 
density distributions $\rho(r)$. These are based on  muonic 
atoms and electron scattering data.  It is known that these experiments 
determine  precisely    $R_{ms}$ -the mean square radii of the charge 
distribution.  However,  for  the problem of  interest  one  needs  higher 
moments of the proton density distribution, and these  are  not known 
very precisely.  Since the atomic  wave 
function behaves as $r^{l}$  one could expect the $r^{2l}$-th moment 
to play the dominant role.  More realistically,  due to the   exponential 
term in the wave function of Eq.(\ref{B2}) it is the   $r^{2l-2}$-th moment 
of the density that gives the main contribution.  According to  Eq.(\ref{18})
we are interested in the  moments of   proton distributions  folded over 
a form-factor  related to the pion exchange.  
Let us now discuss the relation  of these quantities and the  uncertainties 
involved.  To be specific  we study  the best known  nucleus  $^{40}$Ca.   

The relation of   bare  $\langle r^{2l}\rangle $ and folded  
$ \langle R^{2l}\rangle $  
density moments  may  be  obtained  via studies  of  the corresponding 
Fourier transforms as given in Ref.\cite{wanda}.  For  $l=5$ one obtains   
\begin{equation}
\label{moment}
  \langle R^{10}\rangle  =   \langle r^{10}\rangle + \frac {55} {3} 
(\langle r^{8}\rangle 
\langle u^{2}\rangle +  \langle r^{2}\rangle \langle u^{8}\rangle)  
  +66(\langle r^{6}\rangle \langle u^{4}\rangle
  +  \langle r^{4}\rangle \langle u^{6}\rangle) +  \langle u^{10}\rangle
  \end{equation}
where $ \langle u\rangle $ are the moments of the folding function $F_{fold}$.
Similar relations allow to express the proton density moments 
$ \langle r^{2k}\rangle $ in terms of protonic $ \langle u_{pcharge}\rangle $ 
and nuclear $\langle R_{charge}\rangle $ electromagnetic moments.  The charge
density moments are calculated from the experimental charge densities taken
from the compilations of Refs.~\cite{dens}.  One finds five different
densities:  two sets of 2pF and three sets of 3 parameter Fermi
(3pF) profiles (these 3pF will be defined in the following
paragraph).  Next, we calculate the average proton density moments
generated by those densities. These moments contribute to the folded moment 
 of Eq. (\ref{moment}) with
the weights of 1, 0.71, 0.02, 0.33, 0.10 and 0.001 respectively. Thus
the highest $ \langle r^{10}\rangle $ moment contributes less than half of the whole
folded moment.  On the basis of these five experimental densities one can also
calculate an error (r.m.s deviation) of each proton density moment. Thus the
error of $ \langle r^{8}\rangle$ is 11\% while the error of 
$ \langle r^{10}\rangle$ is as large as 29\%.  A total error of 
$ \langle R^{10}\rangle $ obtained from Eq. (\ref{moment}) becomes 18\%.  The
same analysis done for the folded $ \langle R^{8}\rangle $ moment 
yields an error of 7.8\%.  These large errors offer certain upper limits
since the errors of the $ \langle r^{2k}\rangle$ are correlated and should not
be added independently.

A different procedure to calculate the uncertainty is adopted here: for each
experimental charge distribution we calculate a corresponding 3pF density.
  The parameters of this proton density are
fixed to reproduce the lowest $ \langle R_{charge}^2 \rangle $, 
$ \langle R_{charge}^4\rangle $ and $ \langle R_{charge}^6 \rangle $ 
moments.  Next the
atomic level width is calculated for each folded proton density.  The mean
square error of the average width obtained in this way amounts to 3.4\%,
only. With the same procedure we obtain errors of the other atomic level
widths:
6.4$\%$ in Pb (two  2pF and two Gaussian charge density profiles), 
10$\%$ in Al (three  2pF profiles) and 11$\%$ in Si (two 2pF and two 3pF
profiles). The  overall
uncertainty, weighted by the experimental errors, is
estimated to be about 5$\%$. For each nucleus we chose 
the 
proton density closest to the average
and this density is used for further calculations.  It turns out that in
each case these are the proton densities based on the experimental charge
densities by Fricke et.al~\cite{dens}. 
The parameters $(c,\ a_p,\ w)$ found for the 
3pF ($(1+w(R/c)^2 /(1+e)$) proton density profiles are:
(3.01, 0.493, -0.01) for Al, (3.13, 0.479, 0.016) for Si, 
(3.69, 0.518, -0.09) for Ca and  (6.656, 0.474, -0.01) for Pb.

The $\Sigma$ hyperon in atomic orbit is affected by the complex optical
potential of the nucleus. This changes the atomic wave function in the
nuclear region. Two opposite effects contribute: i) the nuclear absorption
described by Im~$ V^{opt}$ tends to suppress the wave function, ii) the 
Re~$V^{opt}$ is weakly attractive at large distances and tends to enhance the
wave function.  Typical optical potential strengths are described by a
scattering length $b_{0}$. Phenomenological, best fit values are given in
Ref.~\cite{bat97} and the simplest choice is $b_{0} = (.25 +i.15) fm $ which
characterizes an optical potential with the charge density profile. Solving
Schr\"odinger equation for the level width in Pb one finds that the
perturbative calculation underestimates the true width by about~11\%. The
effect of attraction prevails over the repulsive effect due to absorption.
The same exists also within our description of the absorption.  However, 
the
effect is smaller since our absorptive pion exchange potential is longer
ranged that the real potentials.

To account for the multiple scattering effect we express Eq.(\ref{18}) as
the perturbative result for a folded potential as indicated by
Eq.(\ref{2a}). A shift of an argument by $ a {\bf u} $ is necessary to
perform this. Next, such a potential is supplemented by the real optical
potential discussed above.  The level widths obtained from the Shr\"odinger
equation are larger than the perturbative level widths by 5.8\% in Pb ,
2.9$\%$ in Ca , 3.3$\%$ in Si and 3.1$\%$ in Al. The net result of this
calculation is that the coupling constant $f_{\pi \Lambda \Sigma}^{2}$
extracted from the widths becomes smaller than the constant extracted in a
perturbative way. The difference amounts to 4$\%$. These corrections have
been included in the final results given in Table II.

The optical potential correction procedure is not unique as a number of real
optical potentials exist in the literature~\cite{bat97}.  Different
values of $b_{0}$ induce an additional 1$\%$ uncertainty.  This we lump
together with the uncertainty due to the charge distribution . Now , the
total amounts to a 6\% effect.

Thus,  nuclear densities induce the main part of our  
systematic error. The corresponding  error of 0.003 
arises in the $f_{\pi \Lambda \Sigma}^{2}$ coupling constant.

The mixed nucleon density matrix of Eq. (\ref{14}) introduces no serious
uncertainties. As the ranges involved are given by 
$1/\left (2\sqrt{{\bf q}^{*2}+m^{*2}}\right)$ one needs this expression for 
$w<$ 1.5 fm.  The whole effect of $J_{nuc}$ amounts to a reduction of 
$\Gamma$  by less than 5\% and the uncertainties in Eq.(\ref{14}) are 
negligible.

\section{Conclusions}

It has been shown that the upper level widths in $ \Sigma $ 
hyperonic atoms present a good source of information on the
$ f_{\pi \Lambda \Sigma}^{2}$ coupling constants. 
The complications  due to nuclear physics may be kept under 
a reasonable control. The experimental uncertainties still 
present a problem.

Averaging the best fits gives a value of
\begin{equation}
\label{eq:fpilambda}
  f_{\pi \Lambda \Sigma}^{2}/4\pi= 0.048 \pm 0.005\ {\rm (statistic)}
\pm 0.004\ {\rm (systematic)} \ .
\end{equation}
This corresponds to 
 $G^{2}_{\pi \Lambda \Sigma}/4 \pi =  f^{2}_{\pi \Lambda \Sigma}
[ (M_{\Lambda}+M_{\Sigma})/m_{\pi}]^{2}/4 \pi$, i.e. 
\begin{equation}
\label{eq:g2}
G^2_{\pi \Lambda \Sigma}/4 \pi = 13.3 \pm 1.4
\ {\rm (statistic)} \pm 1.1\ {\rm (systematic)}\ .
\end{equation}
This result is obtained
here from the combined analysis of the atomic widths and
the inelastic hyperon scattering cross sections.
This result differs from the numbers used, or obtained in the
analyzes of the scattering data. In particular :
$ f_{\pi \Lambda \Sigma}^{2}/4 \pi = 0.041 $ has been used by
Nijmegen group \cite{swa,sto} while the
J\"ulich group \cite{reu} obtains
$ f_{\pi \Lambda \Sigma}^{2}/4 \pi~=~0.034 $ and/or
$ f_{\pi \Lambda \Sigma}^{2}/4 \pi~=~0.028 $
It turns out that the atomic data favor larger values of the
pion hyperon coupling. Had we used a lower value for the
pion-nucleon coupling constant such as that determined by the Nijmegen
group~\cite{sto} viz., 0.075, instead of the classical value~\cite{ben}, 
0.080, our $ f_{\pi \Lambda \Sigma}^{2}/4 \pi $ would even have been 7\%
larger.

The errors of this determination, summarized in Table III,
 consist of four components: 
i)~the statistical error due to the data uncertainties, 0.005,
 the systematic error due to ii)~uncertainties in the proton 
density distributions, 0.003,  iii)~uncertainties of the energy 
release in the capture process 0.001, iv)~the uncertainty 
due to poor knowledge of the short range hyperon-nucleon 
interactions  estimated to be about 0.003. Altogether, an 
error of 0.007 is obtained.
In order to reduce the large statistical part, both new 
measurements of the $\Sigma$ atomic widths 
and free $ \Sigma^- p \rightarrow  \Lambda n $ scattering data 
would be welcome.

\vspace {1cm}
We thank D. Vautherin for helpful discussions and S. W. is indebted 
to J. Dabrowski for useful comments.
The authors acknowledge support of IN2P3 /INS collaboration agreement  
and  KBN  Grant 2P03B  048 12.


\appendix 
\section {Pion exchange forces}

To make the paper self-contained several useful formulas are collected
in this appendix. These concern the pion exchange in the
$\Sigma N \rightarrow \Lambda N $ conversion.
Due to the $ M_{\Sigma}-M_{\Lambda}$ mass difference this process involves
momentum  and also some energy transfer. It makes the exchange
force to be longer ranged as compared to the  $N-N$ interactions.
The $\Sigma^-$ and $p$ being bound into the nucleus are not on mass-shell
Dirac spinors. To derive the OPE potential we shall nevertheless consider
them as free on mass-shell spinors. The reduction of the
pseudo-vector $\gamma_5 \gamma_{\mu} \partial^{\mu} \phi_{\pi}$
pion-nucleon ($\pi N$) and pion-hyperon ($\pi Y$) couplings leads,
in the momentum representation, to

\begin{equation}
\label{ope1}
  V_\pi({\bf q})= \frac {\sqrt{2} f_{\pi NN}} {m_{\pi}}\ V_N\
                  \frac {f_{\pi \Lambda \Sigma}} {m_{\pi}}\ V_Y\
		  \frac{1} {({\bf  p'}-{\bf p})^2- (\Delta E)^2 +m_\pi^2}
\end{equation}  
where
\begin{equation}
\label{vn}
  V_N= \sqrt{(E_n+M_N) (E_p+M_N)}
  \ \mbox {\boldmath $\sigma$}_N.\left(\frac {{\bf p'}}{E_n+M_N}
      -\frac {{\bf p}}{E_p+M_N}\right)
\end{equation}  
and
\begin{equation}
\label{vsig}
  V_Y= \frac {M_\Sigma + M_\Lambda}{2}
       \sqrt{\frac{(E_\Lambda+M_N) (E_\Sigma+M_N}{M_\Lambda M_\Sigma}}\
  \mbox {\boldmath $\sigma$}_Y.\left(\frac {{\bf p'}}{E_\Lambda+M_\Lambda}
                     -\frac {{\bf p}}{E_\Sigma+M_\Sigma}\right).
\end{equation}
Here ${\bf p}$ and ${\bf p}'$ denotes the
initial and final C.M. momenta of the reaction (\ref{1}),
$E_n=\sqrt{{\bf p'}^2+M_N^2}$, $E_p=\sqrt{{\bf p}^2+M_N^2}$,
$E_\Lambda=\sqrt{{\bf p'}^2+M_\Lambda^2}$ and
$E_\Sigma=\sqrt{{\bf p}^2+M_\Sigma^2}$.
In the free space $\Delta E = E_n-E_p
= E_\Sigma-E_\Lambda$.
If one neglects ${\bf p'}^2$, ${\bf p}^2$ terms and $\Lambda-\Sigma$ mass
difference, Eq. (\ref{ope1}) with Eqs. (\ref{vn}) and (\ref{vsig}) reduce 
the  corresponding momentum representation OPE to
\begin{equation}
\label{a1}
  V_\pi({\bf q})  = \frac { f^2}{m_{\pi}^2}\ 
	      \frac{(\mbox {\boldmath $\sigma$}_N.{\bf q})
             (\mbox {\boldmath $\sigma$}_Y.{\bf q})}
	       {{\bf  q}^2- (\Delta E)^2 +m_\pi^2}
\end{equation}  
where ${\bf q}={\bf p}' - {\bf p}$ is the momentum transfer.
In the atomic state
$\Delta E = E_{\Sigma N} -E_{\Lambda N}$ 
is the 
energy transferred by mesons. The $\Delta E $  of $\approx  40$~MeV 
at low energies makes an effective pion mass 
$ m^* = \sqrt{ m_\pi^2 -(\Delta E)^2} $ 
of about  
 $ 135$~MeV.
 The coupling constant appropriate for 
$\Sigma^{-} p  \rightarrow \Lambda p $  reaction is   
$ f^{2}= \sqrt{2}\ f_{\pi N N}\ f_{\pi \Lambda \Sigma } $. In principle,  
it is the hyperonic constant $ f_{\pi \Lambda \Sigma } $ which is
to be determined here. In reality, this determination is affected by 
a small uncertainty of the pion nucleon constant  $f_{\pi N N}$  \cite{ben}. 

In the coordinate representation the potential of Eq.(\ref{a1}) corresponds to 
\begin{equation}
\label{a2}
  V_\pi(r)  = \frac { f^2}{m_{\pi}^2}\
	      (\mbox {\boldmath $\sigma$}_N.\mbox{\boldmath $\partial$})
(\mbox {\boldmath $\sigma$}_Y.\mbox {\boldmath $\partial$})
	      \frac{\exp(-m^*\ r)}{4\pi r}.
\end{equation}  
This potential is  strongly singular at the origin, and  this singularity 
has to be removed by a zero in the wave function induced by a repulsive  
core potential. A simpler and more customary way is to remove the 
$\delta ({\bf r})$ like term from expression (\ref{a2}),\cite{gib}.
This is conveniently done 
in momentum representation (\ref{a1}) by the splitting 

\begin{equation}                                          
\label{a3}
  (\mbox {\boldmath $\sigma$}_N.{\bf q})
(\mbox {\boldmath $\sigma$}_Y.{\bf q}) = 
  \frac {1}{3} \left(S_T\ q^2+\mbox {\boldmath $\sigma$}_N.
\mbox {\boldmath $\sigma$}_Y \ q^2\right)
\Rightarrow  \frac {1}{3} \left(S_T\ q^2-\mbox {\boldmath $\sigma$}_N.
\mbox {\boldmath $\sigma$}_Y\ m^{*2}\right)
\end{equation}  
where $ S_T=3(\mbox {\boldmath $\sigma$}_N.{\bf \hat{q}})
(\mbox {\boldmath $\sigma$}_Y.{\bf \hat{q}}) -
\mbox {\boldmath $\sigma$}_N.\mbox{\boldmath  $\sigma$}_Y  $ 
is the tensor force that does not induce 
the $\delta({\bf r})$ singularity. The  step indicated by arrow in 
Eq.(\ref{a3}) describes the subtraction of 
$  \delta({\bf r})\ \mbox {\boldmath $\sigma$}_N .
\mbox{\boldmath $\sigma$}_Y \ f^2/m_{\pi}^2$ 
from the spin-spin potential. For the $\Sigma-\Lambda$ conversion one 
finds $q>2m^* $ which makes the tensor force to be the strongly 
dominant one.

The size of pion source is described by a vertex function
$ F_{\pi BB}= (\lambda^2-m^{*2})/(\lambda^2+q^{2}) $ that induces a change
in the pion potential $ V_\pi(q) \Rightarrow$ 
$V_\pi(q) F_{\pi BB} F_{\pi \Lambda\Sigma}$. The same source size 
parameter $\lambda = 770$~MeV is used for the nucleon and hyperon vertices. 
These vertex functions induce a change in the spatial form of potential  
(\ref{a2})
\begin{equation}                                          
\label{a4}
\frac{\exp(-m^* r)}{r} \Rightarrow \frac{\exp(-m^* r)}{r}-
\left (\frac{\lambda^2-m^{*2}} {2\lambda}
+\frac{1}{r}\right)\ \exp(-\lambda r)
\end{equation}  
This finite source potential is weaker than the point source one. However, 
in atomic states of high $l$ the effect of source size is very small as 
the centrifugal barrier suppresses short distances. 

Keeping only first order corrections in the atomic states
 one has
$E_n \simeq M_N+{\bf p'}^2/(2M_N)$, $E_p \simeq M_N-E_B$,
$E_\Lambda \simeq M_\Lambda+{\bf p'}^2/(2M_\Lambda)$
and
$E_\Sigma \simeq M_\Sigma$.  Eqs. (\ref{vn}) and
(\ref{vsig}) reduce to
$$  V_N \Rightarrow  \left(1-\frac{E_B}{4M_N}
             -\frac{{\bf p'}^2} {8M_N}\right)
	     \mbox {\boldmath $\sigma$}_N.{\bf q}$$
and
$$  V_Y \Rightarrow  \left(1+\frac{M_\Sigma - M_\Lambda}{2(M_\Sigma +
M_\Lambda)}
             -\frac{{\bf p'}^2}{8M_\Lambda}\right)
	     \mbox {\boldmath $\sigma$}_Y.{\bf q}$$
It can then be seen that the coupling $f^2$ of Eq.(\ref{a2})
is renormalized by
\begin{equation}
\label{Nf}
N_f=\left(1+\frac{M_\Sigma - M_\Lambda}{2(M_\Sigma + M_\Lambda)}
      -\frac{E_B}{4M_N}
             -\frac{{\bf p'}^2} {8M_\Lambda} -\frac{{\bf p'}^2} {8M_N}\right)
\end{equation}
$E_B$ and ${\bf p'}^2$ take slight different values depending on the nucleus
considered but on the average $N_f$=1.02. The $f_{\pi \Lambda \Sigma}$ values
quoted in this work have been corrected by this renormalization.
	     
The cross section 
for $ \Sigma^- + p \rightarrow \Lambda + n$ within the Born  one pion 
exchange approximation becomes   
\begin{equation}                                          
\label{cross}
\frac{d \sigma} {d \Omega} = \frac{q_{\Lambda N}} {q_{\Sigma N}}
\left(
\frac{2\mu_{\Lambda N} \sqrt{2} f_{\pi NN} f_{\pi \Lambda\Sigma} }
{4 \pi\ m_{\pi}^{2}} 
\right)^{2}
 F_{\pi BB}^2\ F_{\pi \Lambda\Sigma}^2\
\frac{m^{*4}/3 +2 q^{4}/3 } {(m^{*2}+q^2)^{2}}
\end{equation}
where $ q_{\Lambda N}$($q_{\Sigma N}$) are the final(initial) c.m. momenta.

\section{Atomic density matrix}

The unitarity expression for level widths requires some simple
representation of the mixed atomic density
\begin{equation}
\label{B1}
\rho_{at}({\bf Z},{\bf z}) = \frac{1}{2l+1}\sum_m \overline {\Psi}({\bf Z}+{\bf z})
                                                  \Psi({\bf Z}-{\bf z})
\end{equation}
In the text we have ${\bf Z} = {\bf R} +a\ {\bf u}$ and
${\bf z} = a\ {\bf w}/2$. The $\rho_{at}({\bf Z},{\bf z})$ is evaluated for
small $z/Z$ and the expansion parameter becomes
$\epsilon = z^2/Z^2 \approx (b/2)^2 w^2/R^2$. For Pb atom this quantity is
very small, $\epsilon < 0.002$, but the  atomic  wave function change 
rapidly in space and some care is necessary.

The states of interest are the circular orbits and
\begin{equation}
\label{B2}
\Psi_{nlm}({\bf Z})= N_{nl}\ Y_{lm}(\hat Z)\ Z^l\ \exp(-\frac{Z} {nB})
\end{equation}
where $B$ is the Bohr radius and $N_{nl}$ is a normalization.
To perform the expansion let us notice that
$({\bf Z}+{\bf z}).({\bf Z}-{\bf z}) = Z^2\ (1-\epsilon)$,
$|{\bf Z}+{\bf z}||{\bf Z}-{\bf z}| = Z^2\ (1+2\epsilon-4\epsilon c^2)^{1/2}$
and
$$\frac{({\bf Z}+{\bf z}).({\bf Z}-{\bf z})}{|{\bf Z}+{\bf z}||{\bf Z}-{\bf
z}|}=x \sim 1-2\epsilon+2\epsilon c^2$$
where $c=\cos(\hat Z \hat z)$. The most important correction 
in Eq. (\ref{B1}) is
due to spherical harmonics as the hyperon momenta are predominantly
tangential to the nuclear surface. One has
\begin{equation}
\label{B3}
\frac{4\pi}{2l+1} \sum Y_{lm}(\widehat {Z+z}) Y_{lm}(\widehat {Z-z}) = P_l(x)
                   \approx P_l(1-2\epsilon+2\epsilon c^2)
		   \approx 1-l(l+1)\epsilon(1-c^2)
\end{equation}
where we use $P_l'(1)=l(l+1)/2$. The other pieces of atomic wave
functions expanded in a similar way produce much smaller corrections. Thus
$|{\bf Z}+{\bf z}|^l|{\bf Z}-{\bf z}|^l  \approx Z^{2l}\ (1+l\epsilon(1-2c^2))$
and
$$\exp\left(-\frac{|{\bf Z}+{\bf z}|+|{\bf Z}-{\bf z}|}{nB}\right) \approx
\exp\left (-\frac{-2Z}{nB}\right )\ \left (1+\frac{\epsilon}{nB} \right ).$$
Altogether one obtains
$$\rho_{at}({\bf Z},{\bf z})  \approx \rho_{at}({\bf Z},0)\ 
                             P_l(1-2\epsilon+2\epsilon c^2)\ 
			     (1+l\epsilon(1-2c^2))\ \left (1+\frac{\epsilon}{nB}\right )$$
\begin{equation}
\label{B4}
             \approx \rho_{at}({\bf Z},0)\ P_l(1-4\epsilon/3)\
	      (1+l\epsilon/3)\ \left (1+\frac{\epsilon}{nB}\right )
\end{equation}
where in the last line an average value of $c^2=1/3$ was used. In practice
it is only the $P_l(1-4\epsilon/3)$ term that is the important one giving a
few percent correction to the atomic density $\rho_{at}({\bf Z},0)$.



\pagebreak

\begin{table}[htb]
\caption{Experimental and calculated results for upper level widths, 
$\Gamma$(eV), 
in circular orbits. $l$ is the level angular momentum. }

\begin{center}
\begin{tabular}{ccccccc}
	  &Atom & Ref.             & $l$ & $\Gamma_{\rm experiment}$(eV) &
$\Gamma_{\rm calculated}$(eV) &\\ \hline
 & $^{27}$Al&\cite{bat78}   & 4   & 0.24(6)   &   0.21 &  \\
 & $^{28}$Si&\cite{bat78}   & 4   & 0.41(10)  &   0.51 &  \\
 & $^{40}$Ca&\cite{bac}     & 5   & 0.41(22)  &    0.17 & \\
 & $^{208}$Pb&\cite{pow}    & 9   &  17(3)   &    15 &\\
\end{tabular}
\end{center}
\end{table}

\begin{table}[htb]
\caption{Summary of the best fits to the atomic and scattering data. }
 
\begin{center}
\begin{tabular}{ccccccc}
   Model& A$(\pi)$ &B($\pi + V_1(2\pi)$)& C($\pi + V_T(\rho)$)&D($\pi +
K$)&E($\pi + \rho$)&
   F($\pi + K + \rho $)\\
\hline
atomic $\chi^2$/ data  &2.42/4  &2.42/4  &2.33/4  &2.31/4  &2.43/4 & 2.28/4 \\
full   $\chi^2/data$   &5.90/12 &5.84/12 &5.74/12 &5.67/12 &5.89/12& 5.58/12 \\
 $f_b^2$               & 0      &-0.09    &2.4     &-0.19   &0.4    & -0.20
($K$); 1.7 ($\rho$)   \\  $f_{\pi \Lambda
\Sigma}^2$&0.047(4)&0.045(4)&0.050(4)&0.051(5)&0.047(5)& 0.051(5) \\
\end{tabular}
\end{center}
\end{table}

\begin{table}[htb]
\caption{Summary of errors in the determination of 
$f^2_{\pi \Lambda \Sigma}$. }
 
\begin{center}
\begin{tabular}{ccccccc}
  & Source & Energy & Proton density &
Background &Statistics&\\
   &of error&  release& distribution&
 & &\\
\hline
&Error &0.001 &0.003 &0.003 &0.005 &\\
\end{tabular}
\end{center}
\end{table}

\pagebreak

\end{document}